# The Drought-Stress Response of a Drought Resistant Impatiens Processed in PEG-6000 Solution in a Simulation Test

Xue Lan Liao[2,3]   Guo Qin Wen[3]   Qing Lin Liu[2]   Xue Yang Li[2]   Meng Xi Wu[2] and Yuan Zhi Pan[2,4]

Author Affiliations

[2]College of Landscape Architecture, Sichuan Agricultural University, Chengdu, Sichuan,China

[3]Key Laboratory of Southwest China Wildlife Resources Conservation (China West Normal University), Ministry of Education, Nanchong, Sichuan 637002, China

**Abstract**:

**Premise of the Study**: Impatiens is a commonly seen garden flower, renowned for its strong adaptability and long history of cultivation. However, seldom has any research touched on its physiological resistance mechanism. In this experiment, the impatiens is selected from those which experienced aerospace mutation and thereafter 12 years of cultivation and breeding. Therefore, it is superior to the non-mutagenized impatiens in terms of drought resistance and demonstrates tremendous differences from the normal impatiens in physiology, which intrigues scholars to search for the underlying reasons .

**Methods:** By reference to *Impatiens balsamina* L,this experiment uses mutagenized impatiens seeds, processed by PEG-6000 in different solution concentration, to measure the germination rate of impatiens, its relative enzymatic activity and expression differences between gene SoS2 and gene RD29b in the drought lower reaches.

**Key results:** Under simulated drought stress , there is no distinct difference between the mutagenized impatiens and the normal impatiens in terms of germination rate. But by measuring the root tillers and the length, the relative enzymatic activity, MDA, and the expression differences between gene SoS2 and gene RD29b in the drought lower reaches, it is verified that the mutagenized impatiens has more advantages than the normal impatiens, and it Can further cultivate become drought resistance varieties impatiens.

**Conclusions**: In this experiment,which is a positive mutation, the mutagenized impatiens improves drought resistance through radiative mutation. The so obtained impatiens is more pleasing in sight in terms of color and shape and has higher application value in garden virescence.

**Key words**: mutagenized impatiens, germination of seed, adversity，PEG-6000 ，SOD POD, gene expression

This work was funded by the 13th Five-Year Plan for Economic and Social Development of the People's Republic of China---breeding research projects in sichuan province.

[4]Author for correspondence: Pan (scpyzls@163.com), phone: 86-13880886911

Impatiens is the biggest flower genus in Balsamine family.There are about 900 species around the world, distributed mainly in tropical and subtropical mountain areas and sparsely in north temperate zone. The reported Impatiens in China has about 200 species, growing mainly in Southwest provinces. Impatiens, also known as Henna,Henna dye and peach flower, has a scientific name in Latin:Impatiens balsamina L. . Impatiens is an annual herbaceous flower widely planted in gardens for appreciation with long florescence, strong adaptability, and marvelous colors. It is so named due to the observation that its head,wing,tail and foot outstand like the phoenix bird and therefor plays some role in China's flower culture. It is also named "camphire" and "daughter flower" due to the fact that it contains natural brick red pigment which can be used to paint fingernails. Because of the combination of ornamental and practical values, impatiens is much favored by Chinese intellectuals. As a result, it finds its frequent appearance in their literary works, which gives rise to the richly historied and highly connotative impatiens culture(CHAI Jihong,2012)

In the past, plant biosystematists, domestic or foreign, conducted little research on Chinese Impatiens and the relative documents are in lack. A further study is needed on its constituents, growing mechanism and application value to fill the vacancy. At present, there is a large quantity of studies on the pharmacological constituents and operating mechanism of impatiens at home and abroad while little is on its physiological mechanism of resistance. Since the 60s of the last century, with the human exploration of space, opening the human using special environmental conditions of plant space mutagenesis, through breeding and many new varieties of crops and ornamental flowers. Research shows that after space mutagenesis of plant material, shape, enzyme activity, flowering, etc will change(LIU Min et al.,2003; JIANG Yifan et al.,2007; CHEN Yu et al.,2013; ZHANG Yongfu et al.,2015).

The impatiens balsamina L. used in this experiment is the aerospace mutated No. 2 impatiens in China West Normal University, descending from No.2 seed of the SP14th generation of the mutated impatiens balsamina L(TANG Zesheng,2007). The aerospace mutated No. 2 impatiens in China West Normal University appears red. The base of its main stem is red and on its top stands a rose-shaped large flower without pistill and stamen. Flowers formulated K1-3C5_15A5G1 on the rest part mainly have compound petals, the maximum of which can reach 20. The plant is hemispherical or almost spheroidal and is plastic and resistant to diseases. The No. 2 impatiens in China West Normal University has a long florescence of around three months. It usually blooms from June to September in the summer season of high temperature, when other flowers are withered. Pre-research staff have already conducted relative work to its genetic character and chromosome changes during meiosis period, but deeper exploration into the physiological regulation of impatiens in adversity and its affecting factors is need(TANG Zesheng et al.,2007; TANG Zesheng et al.,2006; TANG Zesheng et al.,2005;FAN Zengli et al.,2014 ).

Therefore,in order to determine the superiority of the mutanized balsam, this experiment choosed to measure the germination of bslsam seeds,the enzymatic activity related to resistance and the expression of genes related to drought threat under drought stress.

## Material and method

**Material**

The material of impatiens, after its space journey through "Shenzhou IV" spaceship, the property of which tends to be more stable and was believed to be safely hereditary after cultivation and observation from 2011 to 2016 . The other type of impatiens also used in this experiment was the non-mutagenized one of the same genus.

**Stress procession of seeds**

According to the preliminary test results of three times, the concentration gradient, light intensity, humidity and temperature conditions were identified. Impatiens seeds were randomly chosen among those even grained, plump eared and disease free. There were 3 treated groups in the normal impatiens with each containing 3 packs of 30 seeds, so were in the mutagenized impatiens. The seeds, the other side of which was pressed tightly by filter paper, were placed in order on the central part of glass panes . Then one hollow end of this device (filter paper plus glass panes) was dipped into PEG-6000 solution with respective concentration of 0%, 5%, 10% so that solution can diffuse and saturate the seeds on the central part. The opening of the vessel inside which the glass panes were placed was pressed with a cover to prevent water from overevaporating. Then the seeds processed in different solution concentration were respectively put into the artificial growth cabinet MGC-350HP-2 manufactured by Shang Hai Yi Hen Scientific Instrument Ltd. for nurture. Inside the artificial growth cabinet MGC-350HP-2, the distribution of illumination time was 12hrs sunshine duration/ 12hrs darkness duration and the simulated sunlight strength was set 6600lx resembling the outdoor indirect sunlight in daytime; Humidity level was set 75%; Temperature was set 25hrs darkness duration and the  in darkness duration. During the experiment, culture solution was replaced once every 48 hours to minimize the negative effect of evaporation on the PEG-6000 solution concentration.

**Determination of seed growth index**

The germination criterion is determined by the observation that the radicles stretched 2mm out of the outer coat of seeds(YAN Xingfu et al.,2016 ). How the seeds sprout, i.e. the number of shoots and the germination energy, was observed and documented statistically every 48hrs.The calculation of germination rate was completed one week after the seeds cease sprouting. The experiment duration was 23 days. And the germination energy was observed from the 4th day on to the 6th day after sowing. After the completion of this experiment, vernier caliper was utilized to measure the taproot length of shoots and the tiller number was documented.

The experiment lasted for 23 days. At the end, shoots with healthy momentum of growth were selected and measured for their enzymatic activity POD、SOD、CAT related to stress resistance and the concentration of malonaldehyde MDA by kits manufactured by Nan Jing Jian Cheng Biotech Company.

In order to further understand the expression differences of genes related to drought resistance between these two types of impatiens, extraction kits RNA and reverse transcription kits manufactured by TAKARA Company were utilized, along with BIO-RAD SsoFast real-time PCR kits, to conduct quantitative fluorescent analysis. In the research of stress in adversity, kinds of genes encodes a protein that is induced in expression in response to water deprivation such as cold, high-salt, and dessication. The response appears to be via abscisic acid. The promoter region contains two ABA-responsive elements (ABREs) that are required for the dehydration-responsive expression of RD29B and SOS2 as cis-acting elements. Protein is a member of a gene family with other members found plants, animals and fungi(XU Zhihong et al.,1998 ). Therefore, in this experiment,the conservative sequence 18S of eukaryote was selected as internal reference gene. And with reference to relative literature research, sequences of RD29B and SOS2 were designed, as demonstrated in Fig.1, to measure the expression differences of these two genes between the test materials(LUO Xia et al.,2015; YAN Ping et al.,2008).

| Gene (access number) | | Sequence (5´→3´) | Product (bp) |
|---|---|---|---|
| 18S | F | GTAGTCATATGCTTGTCTC | 209 |
| | R | GGCTGCTGGCACCAGACTTGC | |
| RD29B | F | TTCGGCCATATGTCATCGTTCTCTC | 244 |
| | R | ATGCTCCCTTCTCATGATGCTCTTC | |
| SOS2 | F | GGCAAGTACGAGGTTGGTCGCA | 223 |
| | R | TCGAAGGACTCGCCAACACCTCA | |

Table 1    Primers for quantitative real-time PCR

**Data analysis**

　　Excel 2010 、biorad CFX96 programs were employed for data sorting, statistic analysis and drawing. Data in the drawing were expressed in the standard deviation of the mean ±.

**Results**

**Stress effects of PEG-6000 in different concentration on impatiens' seed germination and root tillers**

Effects on the germination rates and germinabilities of the normal impatiens and the aerospace mutagenized impatiens

  Through to the germination rate and germination potential data sorting, *impatiens balsamina* L. is shown in figure 1,it illustrates that there was no distinct difference between the mutagenized impatiens and the normal impatiens in terms of germination rate and the germination rate decreases with the increasing concentration of PEG-6000. Among the differently processed seed groups, germination rate of the normal impatiens noticeably surpasses that of the aerospace mutagenized impatiens. The seedlings of the normal impatiens grow evenly and 10 days later the germinability ceased developing. With the increasing concentration of solution PEG-6000, the germinability of all the processed groups of the normal impatiens decreases. By comparison, the aerospace mutagenized impatiens sprouts more slowly, but for a longer time. It ceased sprouting 16 days later. The overall germination rate was close to that of the normal impatiens. And the germinability also decreased with the increase of solution concentration.

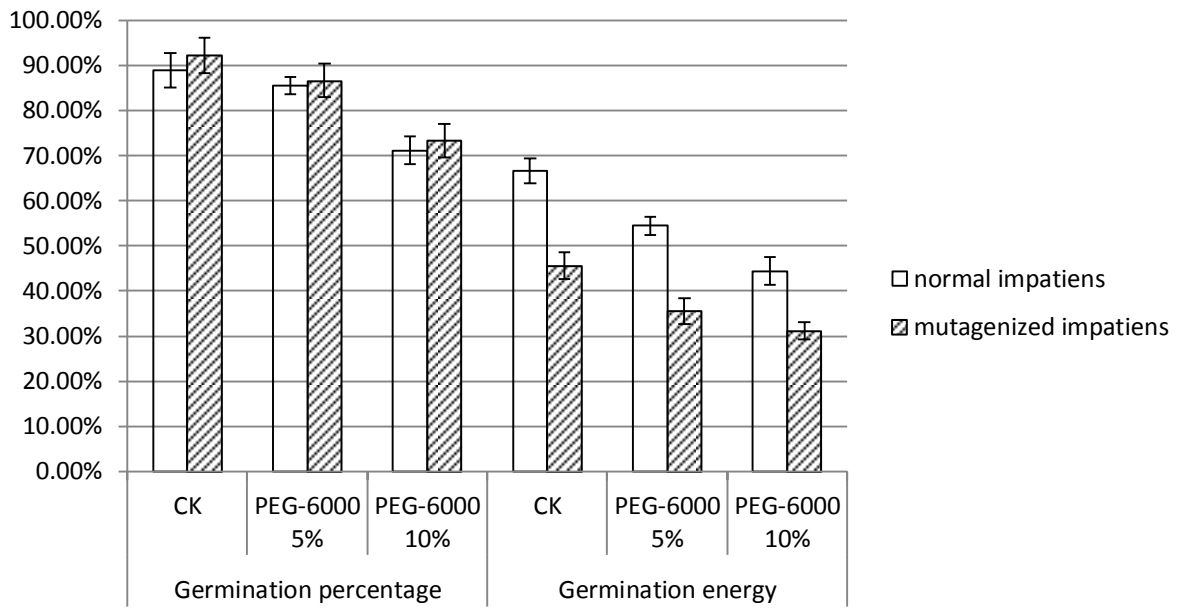

Fig.1 The germination rate and gerninability of both types of impatiens under PEG-6000 stress

Effects on the root tiller number and the taproot length of the normal impatiens and the aerospace mutagenized impatiens

Within the CK groups, the mutagenized impatiens is superior to the normal impatiens in terms of the number of root tillers and the length of taproot. Under the simulated PEG-6000 drought stress, the taproot of the mutagenized impatiens stretches noticeably shorter than that of the normal impatiens, the distinction of which is illustrated in Fig.2. There was no significant correlation between the different concentration of processing solution and the changes of the number of normal impatiens' root tillers and the length of taproot. In terms of the number of root tillers and the length of taproot of the mutagenized impatiens , the experimental groups processed in 5% and 10% concentration were superior to that of the CK group, but the data decline with the intensification of drought stress.

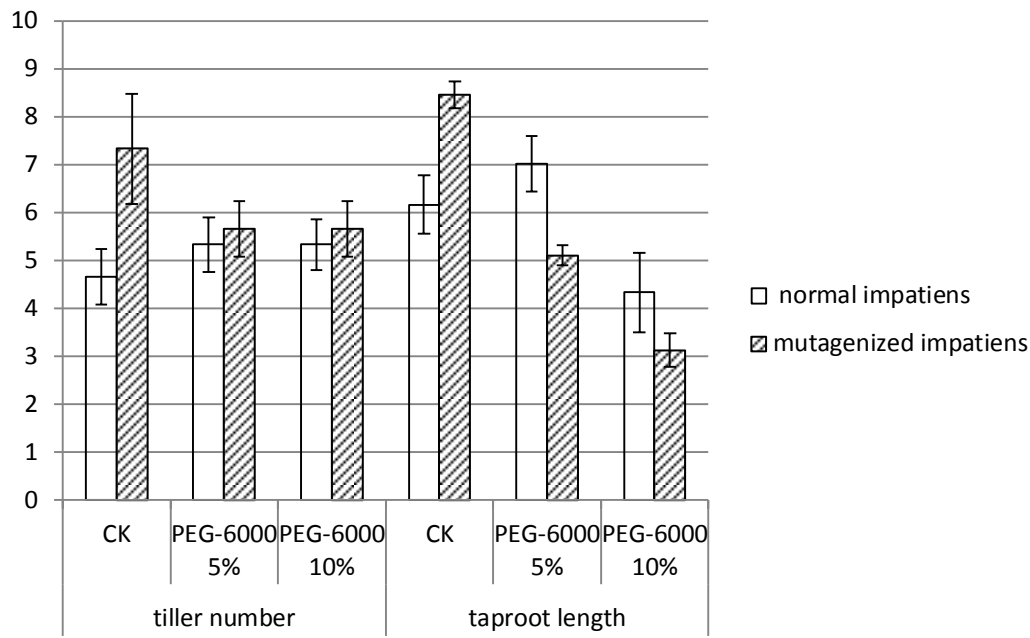

Fig.2 The number of root tillers and the length of taproot under PEG-6000 stress (measurement of root tillers: branch; measurement of the length of taproot: centimeter)

**Effects of PEG-6000 in different concentration on the normal impatiens and the mutagenized impatiens in terms of their enzymatic activity related to stress resistance at the cotyledon stage**

Effects on the content of relative enzymatic activity POD,SOD,CAT

In adversity, the activity of protective enzyme would change to degrade various active oxidation substance generated inside plants. In the simulated PEG-6000 drought stress, the enzymatic activity POD.SOD.CAT of both types of impatiens have undergone some changes, as illustrated in Fig.3 and Fig.4. With regard to the normal impatiens and the mutagenized impatiens, the POD of both experimental groups was higher in comparison to that of CK groups, and the POD of the experimental group processed in 5% solution concentration was observably higher than that of the experimental group processed in 10% solution concentration. In CK groups, there was no distinct difference of the POD activity between both types of impatiens. However in the other two treated groups, the POD of the normal impatiens is slightly higher than that of the mutagenized impatiens.

There was a high indication of SOD in the experimental groups of both types of impatiens. In the procession of 5% solution concentration, the content of SOD in both types of impatiens greatly increases in comparison to that of the CK groups, the phenomenon of which was more observable in the mutagenized impatiens than in the normal impatiens. In each experimental group, the content of SOD in the normal impatiens is inclined to increase in low concentration and decreases significantly in high concentration. The content of SOD in the mutagenized impatiens significantly increases in low concentration and decreases a little in high concentration, but still remained higher that of the CK group.

The CAT was demonstrated differently in the experimental groups of these two types of impatiens. In the normal impatiens, the CAT of the experimental group processed in 5% solution concentration is noticeably higher than that of the CK group. However, with the increase of concentration, CAT of the experimental group processed in 10% concentration decreases sharply and is evidently lower than that of the CK group. With regard to the mutagenized impatiens, the CAT of the experimental groups processed in 5% and 10% concentration both was notably lower than that of the CK group.

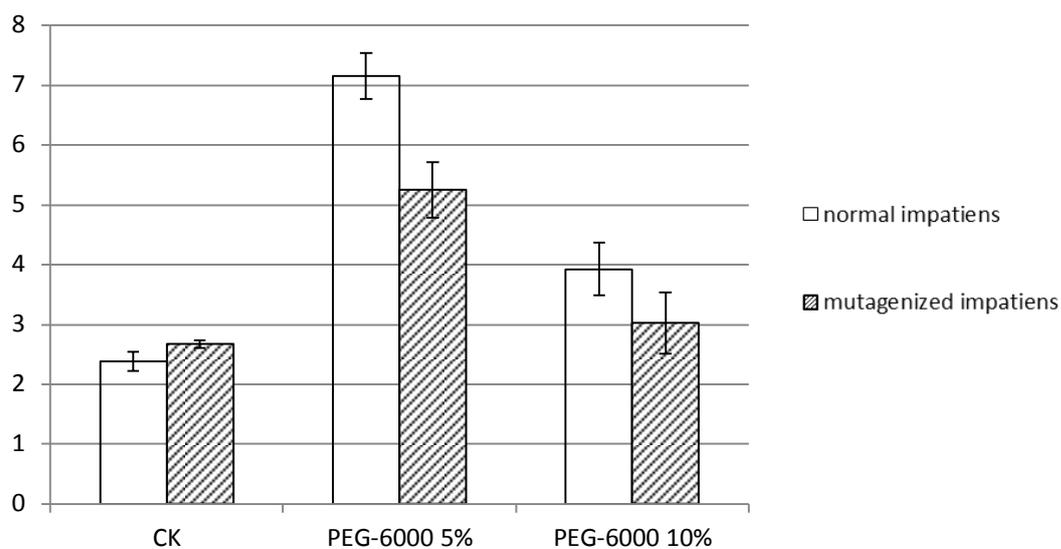

Fig.3　POD activity of two types of impatiens under PEG-6000 stress

（measurement： U/mgprot）

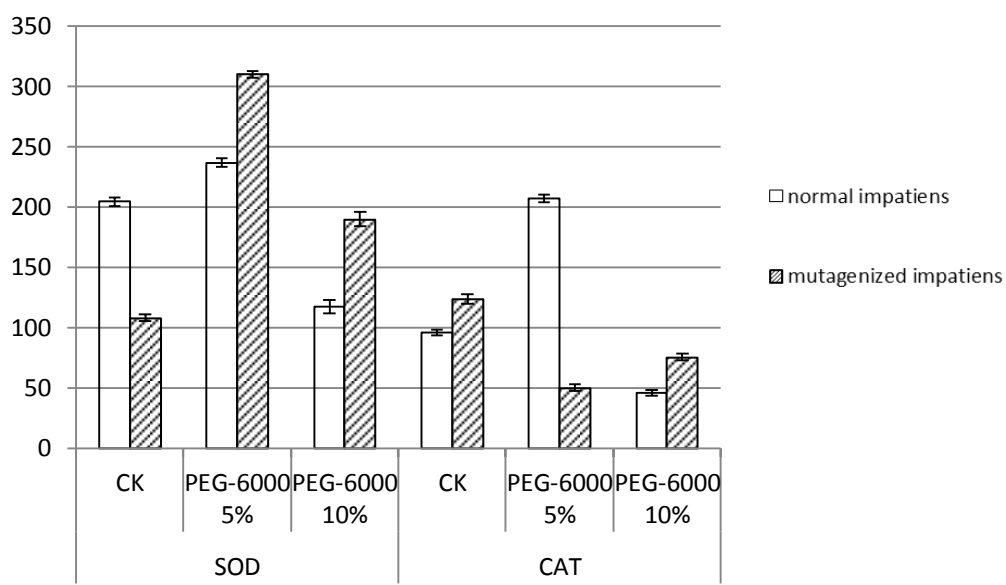

Fig.4 SOD and CAT activity of two types of impatiens under PEG-6000 stress

(measurement:U/mgprot）

Effects on the content of　Malondialdehyde(MDA)

　　MDA mirrors to what extent the plant cell membrane is overoxidated and damaged in adversity, as illustrated in Fig.5. There was a distinct difference of MDA content between the experimental groups of the normal impatiens and the mutagenized impatiens with the former exceeding the latter. No great difference was observed of MDA content between the experimental group of normal impatiens processed in concentration 5% and the CK group. But in the experimental group processed in concentration 10%, the MDA content obviously increases. In each experimental group of the mutagenized impatiens, no significant changes were detected in MDA content.

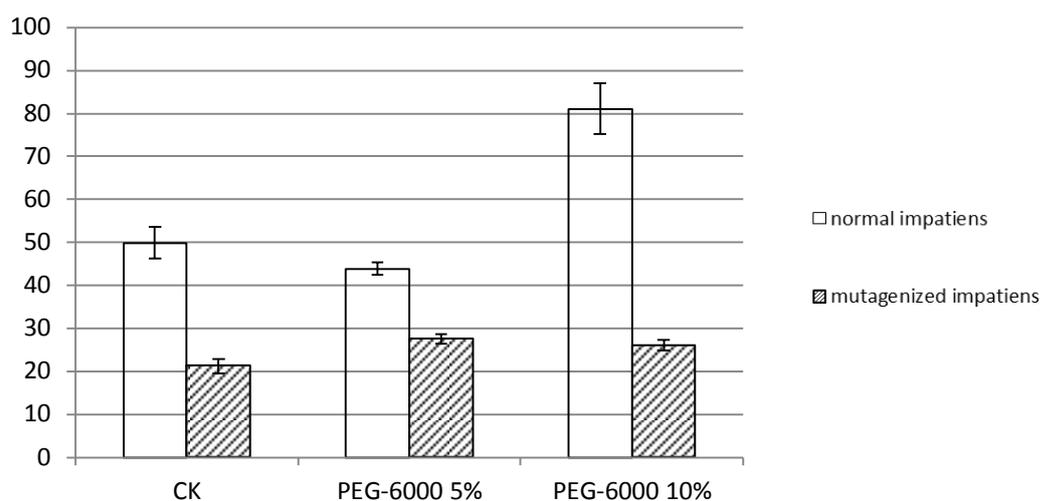

Fig.5 The MDA content of two types of impatiens under PEG-6000 stress

(measurement: nmol/g）

**Effect of PEG-6000 in different concentration on the expression of SOS2 and RD29B genes related to drought resistance in the normal impatiens and the mutagenized impatiens at the cotyledon stage**

Under drought stress, the quantity and changes of the expression of genes relate to drought resistance can index the strength of drought resistance of a plant at the molecular level. In this experiment, quantitative fluorescent examination of genes SOS2 and RD29B in the drought lower reaches were conducted to detect the changes of gene expression of both types of impatiens under drought stress, as illustrated in Fig.6. With regard to the normal impatiens and the mutagenized impatiens, SOS2 expression of both CK groups was higher than that of the experimental groups. Within the CK groups, SOS2 expression of the normal impatiens was apparently higher that of the mutagenized impatiens. And under drought stress, the SOS2 expression of both types of impatiens declines a little but with subtle differences in index. With regard to the normal impatiens, SOS2 expression of the CK group was remarkably higher than that of the experimental groups. The same was true of the mutagenized balsam, but the differences between the CK group and the experimental groups were not so obvious compared with the normal impatiens.

  Within the CK groups, RD29b expression of the normal impatiens was notably higher than that of the mutagenized impatiens. The RD29B expression of the normal impatiens under drought stress was noticeably higher than that of the CK group, but only slightly lower than that of the mutagenized impatiens. The RD29b expression of the mutagenized impatiens under drought stress was slightly higher than that of the CK group.

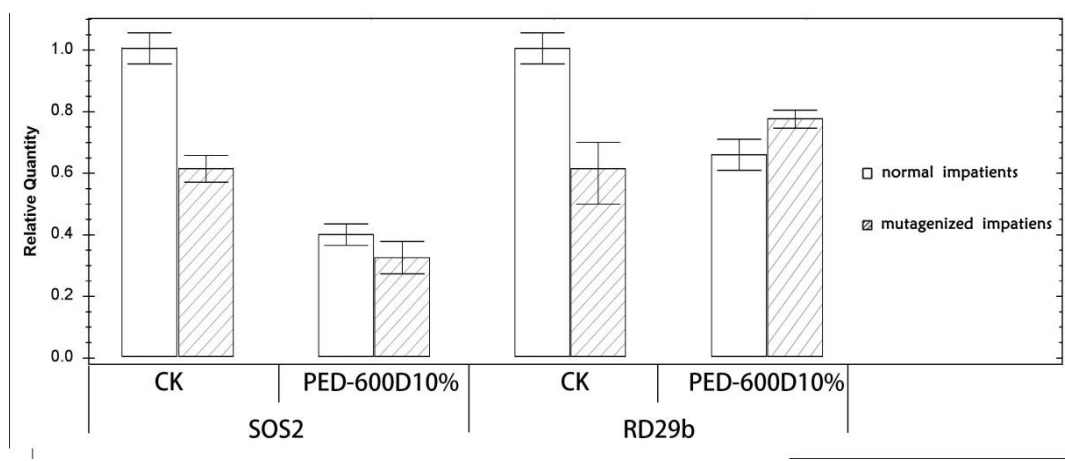

Fig.6 The expression of SOS2 and RD29B genes of two types of impatiens under PEG-6000 stress

## Discussion

In the seed germination experiment, through the observation and analysis of impatiens seed germination rate, germinability, enzymatic activity, and genes related to drought resistance, it was found that the mutagenized impatiens was bestowed with stronger drought resistance, which was consistent with the field observation.

Under the stress of PEG - 6000, impatiens seed germination rate, germinability, taproot length, and tiller number show growth inhibition: the normal germination of seeds is inhibited in high concentration,which conforms to the discovery in the experiment conducted by HUANG Dong et al(HUANG Dong et al.,2015; YAN Xingfu et al.,2016; YANG Xudong et al.,2016 ). However, in comparison to the normal impatiens, the mutagenized impatiens demonstrates stronger sensitivity under stress, which is indicated by lower germination rate, shorter taproot, etc..

In order to discover the operating mechanism that motivates the mutagenized impatiens to demonstrate stronger sensitivity than the normal impatiens , the stress related enzymes are analyzed.

In the measurement of POD, SOD and CAT, except for the content of POD which is lower in the mutagenized impatiens than in the normal impatiens, the content of SOD and CAT in the mutagenized impatiens is higher than that in the normal impatiens in adversity. Meanwhile, the membrane lipid peroxide MDA in examination demonstrates no significant change in experimental groups processed in 5% and 10% concentration and the control group. However, the MDA contained in the experimental groups of the normal impatiens processed in 10% concentration increases significantly, which indicated the normal impatiens deteriorates in its ability to scavenge free radicals under drought stress as a result large accumulation of peroxide, Thus the protective enzyme systems of the mutagenized impatiens are superior to that of the normal impatiens, exhibiting a stronger ability to resist drought stress.

Based on the quantity determination of the expression of genes RD29B, SOS2 related to drought resistance, the expression quantity of both these two genes is lower in the experimental groups of normal impatiens under drought stress than in the CK; with regard to the mutagenized impatiens, RD29B of the experimental groups under drought stress increases slightly and the expression quantity of SOS2 declines a little in comparison to those of the CK , but both are higher than those of the normal impatiens under stress.Therefore, from the determination of the expression quantity of these two genes, it is observed that mutagenized impatiens is superior to the normal impatiens in its ability to resist drought stress.

From the above, with regard to the normal impatiens, there is no significant difference between the experimental groups processed in different concentration under drought stress and the CK group in terms of the taproot length and the number of root tillers. The contraction of taproot length after an increase of concentration indicates that the normal impatiens is not sensitive to water molecules within a certain range and tolerant to less severe water shortage. And in severe drought, taproot growth is restrained. However, through the great advantages of the number of root tillers and the taproot length displayed by the CK group, the mutagenized impatiens suggests good growth condition when water and fertilizer are sufficient. But under drought stress, the number of root tillers decreases and the taproot length is contracted significantly, which suggests that the mutagenized impatiens is extremely sensitive to water molecules. The reduction of root systems also affects the growth of the part of a plant above the ground, slowing down the whole plant growth and delaying the florescence. Except that the plant shrinks in conformation, its growth and external appearance are normal. There are no great differences between the experimental groups in drought and the control group in terms of the membrane lipid peroxide MDA. This is also true between the experimental groups processed in different concentration under drought stress. The above suggests that the mutagenized impatiens is less susceptible to drought. Its morphological change is a favorable adaptation to drought (LI Yun et al.,2016; Zeng Y J al.,2010; Rice K J al.,2001). Compared with the normal impatiens with high accumulation of MDA when processed in 10% PEG-6000 concentration, the mutagenized impatiens is superior in its responsive sensitivity to water molecules and adjustability to adversity through slow growth and shrinkage in size. In summary, the emergency response of the mutagenized impatiens under drought stress is characterized by low growth rate, increase of enzymatic activity, removal of free radicals, reduction in the accumulation of MDA, and increase of expression quantity of genes related to drought-resistance--the inside out favorable adaptation to the environment.

In this study, performance differences are analyzed between two types of impatiens processed in different PEG-6000 concentration under simulated drought stress in terms of their conformation, enzymatic activity and gene expression. It is discovered that the mutagenized impatiens is superior to the normal impatiens in its ability to resist drought. A follow-up study is desired to further probe the causes to the differences of SOD enzymatic activity and the relations between genes expression related to SOD regulation. And a deeper exploration to the underlying mechanism of free radical elimination and MDA aaccumulation in the mutagenized impatiens under drought stress is conducive to unveiling the sealed the mystery .